 \documentclass[final,3p,times,twocolumn]{elsarticle}

\usepackage{amssymb}
\usepackage{amsmath}
\usepackage{color}

\journal{Solid State Science}

\begin{document}

\begin{frontmatter}



\title{Spin-Orbit and Strain Induced Modification in Electrical Properties of Monolayer InSb}

\author{Shoeib Babaee Touski}\ead{touski@hut.ac.ir}
\address{Department of Electrical Engineering, Hamedan University of Technology, Hamedan, Iran}

\begin{abstract}
	In this work, the electrical properties of monolayer InSb in the presence of biaxial strain using density functional theory are investigated. Here, we first explore the band structure of InSb with and without spin-orbit coupling (SOC) consideration. The electron and hole effective mass modify with SOC consideration. The electron and hole effective masses lowered two and ten times, respectively. The location of valleys in conduction and valence band for various strains are explored, and the corresponding effective masses are reported. A lower effective mass is obtained for both electron and hole with applying tensile strain, whereas, the bandgap closes for large tensile strain. A numeric fitting has applied to effective mass versus strain, and an equation for every curve is reported. Finally, the work function of this material for different strains is obtained.
\end{abstract}


\end{frontmatter}

\section{Introduction}
Since the birth of graphene in 2004, it has been paid considerable attention in the scientific world toward exploring properties of new two-dimensional (2D) materials \cite{novoselov2004electric,novoselov2007rise}. The 2D materials have been getting intense research since the last decade because of their unique structure and outstanding electronic, optical, and transport properties \cite{yoon2011good,neto2011new}, which have potential for future applications. Also, 2D materials with high specific surface areas and no dangling bond on the surface give advantages to the bulk.

The III-V compound materials have been known as semiconductors with high mobility and direct bandgap. GaAs and InSb, two members of this family, have shown colossal capability in photoelectric industries. The bulk GaAs and InSb widely used in industry because of their unique optoelectronic properties, high mechanical stability, and direct bandgap  \cite{yoon2010gaas,burstein1954anomalous,vurgaftman2001band,wang2013electronic}.
The indium-based compounds demonstrate materials with small and direct bandgap. Due to their narrow bandgap, they proposed suitable applications for free-space communications, infrared imaging systems, and gas phase-detection systems \cite{hachelafi2009,kirby2002linear}. The high electronic mobility (78,000 $cm^2/Vs$) of InSb points to it as a potential candidate in optoelectronic materials \cite{rode1971electron}. The bulk InSb with a narrow bandgap (0.18 eV at room temperature) can be used in thermal imaging, Schottky diodes, and infrared detectors. The InSb nanosheets display high electron mobility and ambipolar behavior.  InSb nanowires demonstrate high electron mobility as $2.5\times 10^4 cm^2/Vs$ that is affected by large spin-orbit coupling \cite{gul2015towards,jalil2019new}.

Furthermore, a strong interest has recently emerged in group III-V two dimensional and these monolayers have both graphene-like planners and buckled structure \cite{dong2010band,zberecki2012emergence,xu2013stacking,bahuguna2016electric,jalilian2016tuning,zhao2016tuning,jiang2017phonon,zhao2015driving,raeisi2019modulated}. Based on density functional theory (DFT) calculations, \cite{csahin2009monolayer,zhuang2013computational} the stability and electronic band structure of monolayer honeycomb structure of group III-V binary compound has been studied. Monolayer and multilayer boron-nitride with wide bandgap proposed as a high-quality substrate for other monolayers that demonstrates low effects on the transport in monolayer \cite{dean2010boron,babaee2013substrate,touski2020comparative}. Some of these compounds demonstrate a topological insulator phase in mono and bilayer due to their strong spin-orbit coupling (SOC) \cite{yao2015predicted,zhao2015driving}. Bi and Sb atoms from group V, and In and Tl from group III posses high SOC where topological insulator in III-V compounds are based on these atoms \cite{lu2018robust,crisostomo2015robust,zhang2016functionalized}. InSb contains two heavy atoms (In and Sb), and one can expect SOC can highly be modify the electrical properties. Strain and functionalize with halogen atoms or methyl can drive these materials in the quantum spin hall effect phase \cite{li2016robust,li2017gallium}.

In this paper, the electronic properties of monolayer InSb are studied using density functional theory (DFT). The effective masses for electron and hole are calculated with and without SOC consideration. After that, the effective masses and location of valleys are explored for various biaxial strain. Finally, the work function is investigated in the different strains.

\section{Computational details}
Our results are based on density functional theory calculations as implanted in the Quantum ESPRESSO package \cite{giannozzi2009quantum,giannozzi2017advanced}.
Projector augmented plane waves (PAW) pseudopotentials within the PerdewBurke-Ernzerhof (PBE) parametrization of the generalized gradient approximation (GGA) were utilized as an exchange-correlation potential \cite{perdew1996generalized}. Electron-ion interactions were described by the norm-conserving pseudopotentials \cite{goedecker1996separable}. In order to include the spin-orbit interaction, fully relativistic approximation was adopted. The energy cutoff of 612 eV for the plane-wave basis expansion was used throughout all calculations. During calculation, a $12\times12$ Monkhorst-Pack k-point grid was used \cite{monkhorst1976special}. We were first performed full geometry optimization until the forces on the atoms are less than 0.001 eV/$\AA$ and the total energy difference lower than $10^{-8}$ eV.

\begin{figure}
	\centering
	\includegraphics[width=1.0\linewidth]{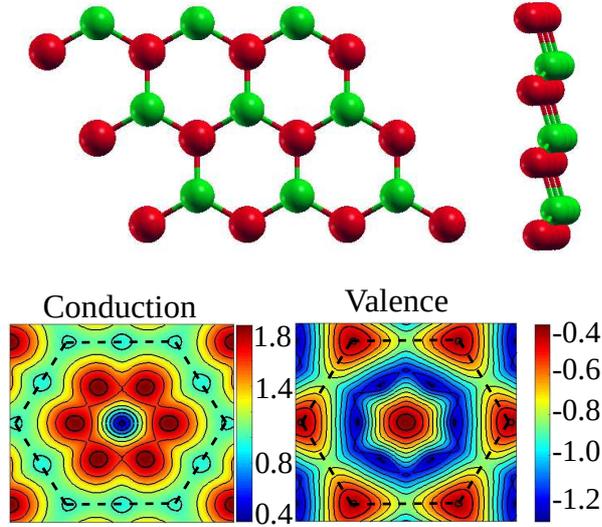}
	\caption{(a) The buckled structure of InSb and (b) the energy of the conduction and valence bands in the first Brillouin zone.}
	\label{fig:fig1}
\end{figure}

The effective masses of electron and hole are calculated by using the following equation\cite{touski2020electrical}:
\begin{equation}
m^*=\hbar^2/\left(\partial^2E/\partial k^2\right)
\end{equation}
Here, $\hbar$ is the reduced Planck constant, E and k are the energy and wave vector of the conduction band minimum (CBM) and valence band maximum. The work function can be obtained the difference between Fermi energy ($E_F$) and vacuum energy ($E_{vac}$) as: $\phi=E_{vac}-E_F$.

\section{Results and discussion}
\begin{figure}
	\centering
	\includegraphics[width=1.0\linewidth]{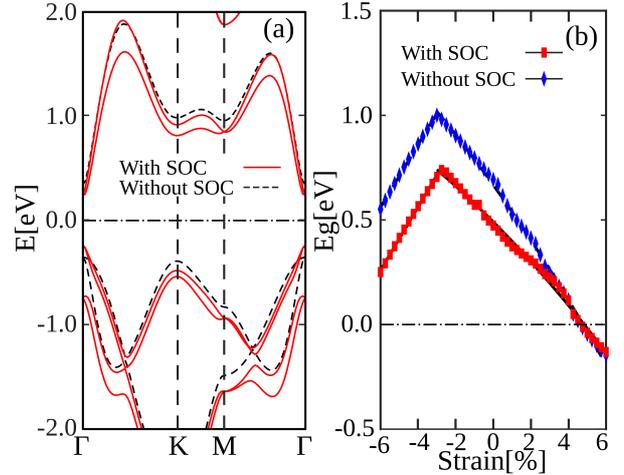}
	\caption{(a) The band structure and (b) the bandgap versus strain with and without consideration of SOC.}
	\label{fig:fig2}
\end{figure}

The structure of InSb with honeycomb configuration is depicted in Fig. \ref{fig:fig1}. The map energy for the first valence and conduction band is plotted in the first  Brillouin zone. As one can observe, $\Gamma$-valley is the conduction band minimum, and K- and M-valleys is much higher than $\Gamma$-valley. On the other hand, both $\Gamma$ and K-valleys contribute to the valence band maximum. The band structure is plotted around the reduced Brillouin zone in a path ($\Gamma$-K-M-$\Gamma$). The band structure of InSb with and without spin-orbit coupling is plotted in Fig. \ref{fig:fig2}(a). The SOC modifies the band structure and changes electrical properties. The CBM and VBM are located in the $\mathrm{\Gamma}$-valley and this material is a direct semiconductor. There is one band at the bottom of the conduction band, whereas, two bands are located at the top of the valence band. Therefore, one band contributes electron effective mass and two bands to hole effective mass.

\begin{table}[t]
	\caption{The effective masses and corresponding energy of the valleys for electron and hole. The effective mass and energy are in the units of $m_0$ and eV, respectively. \label{tab:tab1}}
	\begin{tabular}{p{1.3cm}p{0.8cm}p{1.2cm}p{1.2cm}p{1.3cm}}
		\hline
		\hline
		&    					& w/o SOC & w SOC & Bulk Exp. \cite{vurgaftman2001band} \\
		\hline
		Electron &						&		  &		\\
		& $m^*_{e,\Gamma}$		& 0.0816  & 0.0482 \\
		& $E_{\Gamma}$			& 0.3371  & 0.2392 \\
		& $m^*_e$				& 0.0816  & 0.0482 & 0.014\\
		Hole    &						&		  &		\\
		& $m^*_{h,\Gamma1}$		& ~0.1087  & ~0.0708 \\
		& $m^*_{h,\Gamma2}$		& ~0.5664  & ~0.1355 \\
		& $m^*_{h,K}$			& ~0.6630  & ~0.7070 \\
		& $E_{\Gamma 1}$		& -0.3559 & -0.2395 \\
		& $E_{\Gamma 2}$		& -0.3559 & -0.7434 \\
		& $E_{K}$				& -0.3996 & -0.4787 \\
		& $m^*_h$				& ~0.7978 & ~0.0709 & 0.263 \\
		Band gap				\\
		& Eg					& 0.6930	& 0.4729 & 0.235\\
		\\               
		\hline
		\hline
	\end{tabular}
\end{table}

The bandgap, electron, and hole effective masses are compared for with and without SOC consideration in Table \ref{tab:tab1}. First, the bandgap decreases from 0.6930 to 0.4729 eV with applying SOC. Our band gap without SOC is close 0.68 eV and 0.79 eV in other works \cite{csahin2009monolayer,bi2018thermoelectric} and lower than 1 eV reported by Ref. [\cite{lu2017ab}]. The effective mass of the electron is 0.0816$m_0$, which decreases to 0.0482 $m_0$ with applying SOC. Our electron effective mass is much lower than 0.33 and 0.28 $m_0$ reported in other works\cite{zhuang2013computational, bi2018thermoelectric}. The bulk electron effective mass is 0.014, which is three times lighter than the 2D limit. One can find that two bands at $\Gamma$-point and one band at K-point contribute to VBM and affects the hole effective mass. The effective masses of these three are listed in Table \ref{tab:tab1}. The effective masses at $\Gamma$-point as 0.5664 and 0.1087$m_0$ can be considered as a heavy hole (HH) and light hole (LH). SOC decreases masses for HH and LH to 0.0708 and 0.1355$m_0$, respectively. Both LH and HH in without SOC state crosses at $\Gamma$-point whereas, SOC splits two bands 0.5039 eV. The energy of these effective masses is inserted in the Table. Hole effective mass in the two-dimensional limit can be calculated as:
\begin{align}
\begin{split}
m^*_h=m^*_{h,\Gamma 1}+m^*_{h,\Gamma 2}\exp(-\Delta E_{\Gamma 1,\Gamma 2}/K_BT)\\
+N_Km^*_{h,K}\exp(-\Delta E_{\Gamma 1,K}/K_BT)
\label{Eq:Eg}
\end{split}
\end{align}
where $K_B$ is Boltzmann constant, and $T$ is the temperature in Kelvin. $N_K$ accounts number of K-valley in the first Brillouin zone that $N_K=2$. $\Delta E_{\Gamma 2,\Gamma 1}=E_{\Gamma 1}-E_{\Gamma 2}$ accounts the energy difference between first and second band at $\Gamma$-point. $\Delta E_{\Gamma 1,K}=E_{\Gamma 1}-E_K$ also defines as the energy difference between $\Gamma$ and K valleys in the first band.  The hole effective mass is obtained using the above equation as 0.0709 and 0.7978 $m_0$ for with and without SOC consideration, respectively. The hole effective mass in bulk is 0.263 $m_0$ four times larger than the 2D limit \cite{vurgaftman2001band}. The SOC decreases hole effective mass and makes monolayer InSb as a material with light hole and electron for application as NFET and PFET.

In the following, we apply Bi-axial strain to modify the electrical properties. The compressive and tensile strain is applied, and the resultant structure is completely relaxed, then band structure is obtained, and its electrical properties are extracted. At first, the bandgap as a function of strain is studied with and without SOC, and the results are plotted in Fig. \ref{fig:fig2}. The amount of reduction is higher in the compressive regime, and this indicates SOC is more important in the compressive regime. We observed similar effects in group-III mono-calchagonides \cite{ariapour2020strain}. The bandgap decreases with applying tensile strain, and it goes to zero at 4.5$\%$ tensile strain. For strain larger than this strain,  the CBM is located at $\Gamma$-valley, whereas the VBM at M-valley. The bandgap is zero, but the conduction band and valence band don't cross. InSb is a semi-metal for strain larger than 4.5$\%$ tensile strain. The bandgap first increases with applying compressive strain then decreases for larger compressive strain. The maximum band gap is obtained 0.7386 eV at strain -2.75$\%$. As one can observe, the SOC decreases bandgap in InSb. After this, all calculations are done with consideration of SOC.

\begin{figure}
	\centering
	\includegraphics[width=1.0\linewidth]{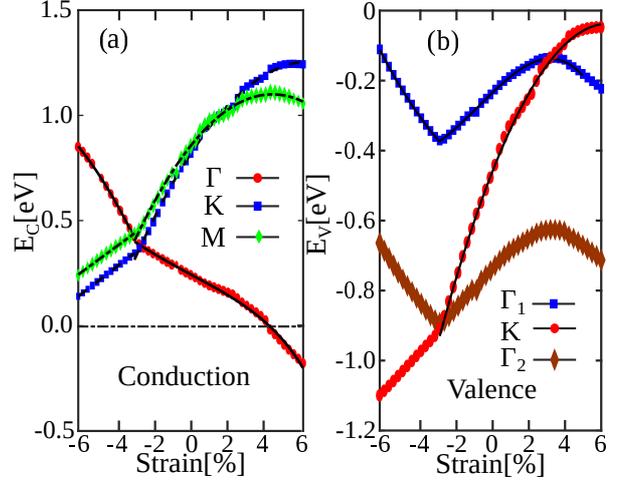}
	\caption{The energy of valleys in (a) the conduction and (b) valence bands for various strain.}
	\label{fig:fig3}
\end{figure}

\begin{align}
\begin{split}
& E_g (-6\leq x \leq -2.75) =  0.1516x+1.17  \\
& E_g(-2.75\leq x \leq 6) = 0.00233x^2-0.09x+0.49
\label{Eq:Eg}
\end{split}
\end{align}

The location of the CBM affects electron transport. This modulates electron density and effective mass. The location of valleys in the conduction band relative to Fermi energy for three valleys ($\Gamma$, K, and M valleys) are plotted in Fig. \ref{fig:fig3}. As one can observe, the CBM is placed in $\Gamma$-valley in tensile strain, and its energy decreases with increasing strain. The energy of valleys increases with applying compressive strain, whereas, K- and M-valleys decrease. At strain -2.75, CBM goes to M-valley; however the energy of K-valley is located close to M-valley. Both M- and K-valley contribute to the conduction band in strain range [-6,-2.75]. In the following, we fit a line to these curves and find the best fitting. We use two ranges of strain for fitting due to breaking in the curves. The first range is [-6,-2.75], and the other is in the interval [-2.75,6]. The relation for fitting is as:
for $-6 \leq x \leq -2.75$
\begin{align}
\begin{split}
E_\Gamma &=-0.145x-0.005308,\\
E_K      &=0.06857x+0.5512,\\
E_M      &=0.06535x+0.6355,
\end{split}
\end{align}
and for $-2.75 \leq x \leq 6$
\begin{align}
\begin{split}
E_\Gamma &=-0.02838e^{x/2.47}+0.2727e^{-x/7.413},\\
E_K      &=-0.01056x^2+0.1363x+0.8176,\\
E_M      &=-0.01287x^2+0.1122x+0.8555.
\end{split}
\end{align}

Two bands at $\Gamma$-point is separated by applying SOC. The energy of two bands at $\Gamma$ point ($\Gamma_1$ and $\Gamma_2$) and K-valley versus strain are plotted in Fig. \ref{fig:fig3}. The second band at $\Gamma$-the point is far from the first one and doesn't contribute to transport. It is obvious that in compressive strain and tensile strain up to 3.25 tensile strain, the VBM is located at $\Gamma$ point. The VBM switches to K-valley for strain larger than 3.25$\%$. The strain is divided into two regimes, first [-6,-2.75] and second [-2.75,6]. The relation for fitted lines is as follow:
for $-6\leq x \leq -2.75$
\begin{align}
\begin{split}
&E_\Gamma = 0.006515x^2-0.024-0.4892
\label{msb}
\end{split}
\end{align}
for $-2.75 \leq x \leq 6$
\begin{align}
\begin{split}
E_\Gamma &=0.001166x^3-0.002163x^2+0.0513x-0.24, \\
E_K      &=-0.01066x^2+0.1365x-0.4741.
\label{Eq:Ee}
\end{split}
\end{align}

\begin{figure}
	\centering
	\includegraphics[width=1.0\linewidth]{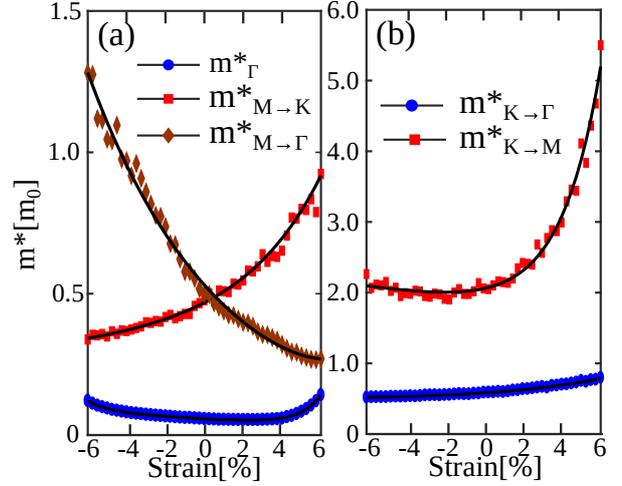}
	\caption{The effective masses of (a) $\Gamma$- and M-valleys, and (b) K-valley in the conduction band.}
	\label{fig:fig4}
\end{figure}

In the first interval [-6,-2.75], the first band at $\Gamma$-point is higher than others, and they have a negligible effect on the valence band. For interval [-2.75,6], two valleys, $\Gamma$, and K have impacts on the hole behavior. Therefore, we do fitting only to $\Gamma$ point in interval [-6,-2.75], whereas to two points, $\Gamma$- and K-valleys in another range.  

The effective mass can be modified by applying strain. The biaxial strain is applied to the sample and electron effective masses for three valleys ($\Gamma$-, K- and M-valleys) are plotted in Fig. \ref{fig:fig4}. $\Gamma$-valley indicates the lowest effective mass. $m^e_\Gamma$ increases with applying compressive strain, whereas, the lowest effective mass can be obtained by tensile strain. The lowest one is 0.045 for 2.5$\%$ tensile strain. The $\Gamma$ effective mass increases after this strain. Two other valleys indicate a higher effective mass than $\Gamma$ valley. The longitudinal effective mass at M valley ($m^e_{M\rightarrow \Gamma}$) increases with increasing strain whereas transverse mass ($m^e_{M\rightarrow K}$)decreases. Both effective masses at K-valley increases with increasing strain. The fitted lines for these five effective masses are:
\begin{align}
\begin{split}
&m^{*e}_\Gamma (-6\leq x \leq 1) = 0.000138e^{-x/1.084}\\
&~~~~~~~~~~~~~~~~~~~~~~+0.057e^{-x/14.9} \\
&m^{*e}_\Gamma (1\leq x \leq 6)=0.0575e^{-x/15.625}\\
&~~~~~~~~~~~~~~~~~~~~~~+0.000364e^{+x/1.065}\\
&m^{*e}_{K\rightarrow\Gamma} = 0.4414e^{-x/114.25}+0.1416e^{+x/6.215}\\
&m^{*e}_{K\rightarrow M} = 1.855e^{-x/51.18}+0.2011e^{+x/2.09}\\
&m^{*e}_{M\rightarrow\Gamma} = 0.3538e^{+x/54.7}+0.1146e^{+x/4.11}\\
&m^{*e}_{M\rightarrow K} = 0.5286e^{-x/6.77}+0.002322e^{+x/1.94}\\
\label{Eq:mse}
\end{split}
\end{align}

\begin{figure}
	\centering
	\includegraphics[width=0.95\linewidth]{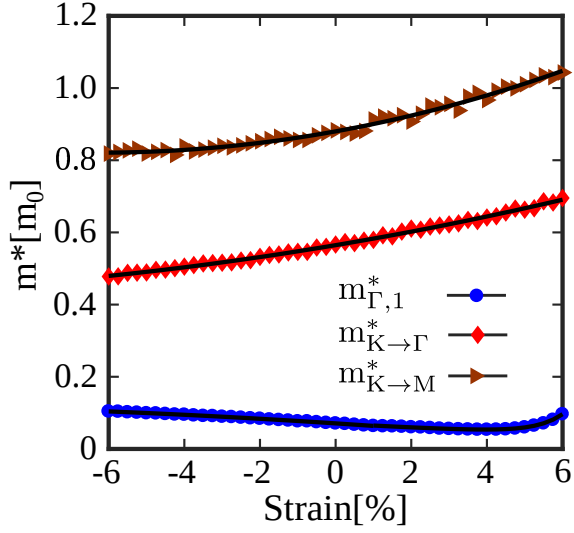}
	\caption{The effective masses of valleys in the first and second bands at the valence band. }
	\label{fig:fig5}
\end{figure}

The effective mass for the valence band is shown in Fig. \ref{fig:fig5}. The hole at $\Gamma$-valley shows a low effective mass but larger than the electron. $m^h_\Gamma$ increases with applying compressive strain, whereas, decreases with tensile strain up to 4$\%$, after that increases. The lowest $\Gamma$ effective mass 0.0534$m_0$ can be obtained at tensile strain 4$\%$. Hole effective mass at K-valley increases with increasing strain. For strain larger than 3.25$\%$ tensile strain, VBM is located at K-point, and InSb indicates a high effective mass for the hole. The fitted line to hole effective masses are:
\begin{align}
\begin{split}
&m^{*h}_\Gamma(-6\leq x \leq 1) =-0.0002174x^2-0.006828x\\
&~~~~~~~~~~~~~~~~~~~~~~~~~~~~~~~+0.07085 \\
&m^{*h}_\Gamma(1\leq x \leq 6) =0.07035e^{-x/12.18}\\
&~~~~~~~~~~~~~~~~~~+1.077\times 10^{-5}e^{x/0.705} \\
&m^{*h}_{K\rightarrow\Gamma} = 0.000565x^2+0.0176x+0.5647\\
&m^{*h}_{K\rightarrow M} = 0.00151x^2+0.01889x+0.88\\
\label{Eq:msh}
\end{split}
\end{align}

\begin{figure}
	\centering
	\includegraphics[width=1.0\linewidth]{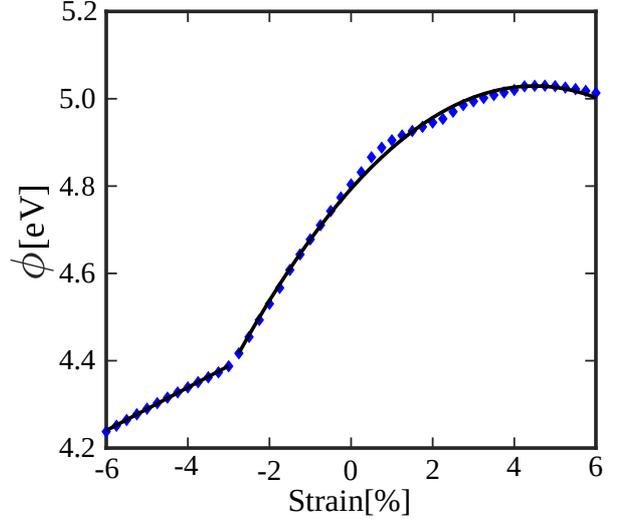}
	\caption{Work-function as a function of strain for InSb monolayer. }
	\label{fig:fig6}
\end{figure}

Another important parameter for studying the electrical property is work function ($\phi$) or electron affinity ($\chi$). The difference between these two parameters is half of the bandgap, $\phi-\chi=E_g/2$. The work function is calculated by the difference between Fermi and vacuum energies. The work function for the unstrained one is 4.68 eV that is close to bulk 4.57eV. The work function is the same for with and without SOC. The work function versus strain is plotted in Fig. \ref{fig:fig6}. $\phi$ behaves linear in interval [-6,-2.75] and parabolic in interval [-2.75,6]. 
\begin{align}
\begin{split}
&\phi(-6\leq x \leq -2.75) = 0.04931x+4.536  \\
&\phi(-2.75\leq x \leq 6) = -0.01156x^2+0.1043x+4.794 
\label{Eq:phi}
\end{split}
\end{align}
The work function can be tuned between 4.2 to 5eV by applying biaxial strain.

\section{Conclusions}
In summary, we investigated the electronic properties of monolayer InSb. We find that the SOC highly declines hole, and electron effective masses to 0.0482 and 0.0711 $m_0$, respectively. In the following, we study the effects of biaxial strain on the electrical properties. The maximum band gap is obtained 0.7386eV at 2.75$\%$ compressive strain, and bandgap closes at 4.5$\%$ tensile strain. The location of hole and electron's valleys versus strain is studied and a numeric fitting has applied to every curve. The effective masses for electron and hole are calculated, and a curve is fitted to them. The lowest electron and hole effective mass occurs when CBM or VBM is located at $\Gamma$-point. At final, the work function increases when strain increases from compressive to tensile.

\section{Conflict of interest}
The authors declare that they have no conflict of interest.

\begin{thebibliography}{10}
	\expandafter\ifx\csname url\endcsname\relax
	\def\url#1{\texttt{#1}}\fi
	\expandafter\ifx\csname urlprefix\endcsname\relax\def\urlprefix{URL }\fi
	\expandafter\ifx\csname href\endcsname\relax
	\def\href#1#2{#2} \def\path#1{#1}\fi
	
	\bibitem{novoselov2004electric}
	K.~S. Novoselov, A.~K. Geim, S.~V. Morozov, D.~Jiang, Y.~Zhang, S.~V. Dubonos,
	I.~V. Grigorieva, A.~A. Firsov, Electric field effect in atomically thin
	carbon films, Science 306~(5696) (2004) 666--669.
	
	\bibitem{novoselov2007rise}
	K.~S. Novoselov, A.~Geim, The rise of graphene, Nat. Mater 6~(3) (2007)
	183--191.
	
	\bibitem{yoon2011good}
	Y.~Yoon, K.~Ganapathi, S.~Salahuddin, How good can monolayer mos2 transistors
	be?, Nano Lett. 11~(9) (2011) 3768--3773.
	
	\bibitem{neto2011new}
	A.~C. Neto, K.~Novoselov, New directions in science and technology:
	two-dimensional crystals, Rep. Prog. Phys. 74~(8) (2011) 082501.
	
	\bibitem{yoon2010gaas}
	J.~Yoon, S.~Jo, I.~S. Chun, I.~Jung, H.-S. Kim, M.~Meitl, E.~Menard, X.~Li,
	J.~J. Coleman, U.~Paik, et~al., Gaas photovoltaics and optoelectronics using
	releasable multilayer epitaxial assemblies, Nature 465~(7296) (2010)
	329--333.
	
	\bibitem{burstein1954anomalous}
	E.~Burstein, Anomalous optical absorption limit in insb, Phys. Rev. 93~(3)
	(1954) 632.
	
	\bibitem{vurgaftman2001band}
	I.~Vurgaftman, J.~{\'a}. Meyer, L.~{\'a}. Ram-Mohan, Band parameters for iii--v
	compound semiconductors and their alloys, J. Appl. Phys. 89~(11) (2001)
	5815--5875.
	
	\bibitem{wang2013electronic}
	Y.~Wang, H.~Yin, R.~Cao, F.~Zahid, Y.~Zhu, L.~Liu, J.~Wang, H.~Guo, Electronic
	structure of iii-v zinc-blende semiconductors from first principles, Phys.
	Rev. B 87~(23) (2013) 235203.
	
	\bibitem{hachelafi2009}
	K.~Hachelafi, B.~Amrani, F.~E.~H. Hassan, S.~Hiadsi, Theoretical study of inas,
	insb and their alloys inasxsb1-x, ACM Physics.
	
	\bibitem{kirby2002linear}
	B.~J. Kirby, R.~K. Hanson, Linear excitation schemes for ir planar-induced
	fluorescence imaging of co and co 2, Appl. Opt. 41~(6) (2002) 1190--1201.
	
	\bibitem{rode1971electron}
	D.~Rode, Electron transport in insb, inas, and inp, Phys. Rev. B 3~(10) (1971)
	3287.
	
	\bibitem{gul2015towards}
	{\"O}.~G{\"u}l, D.~J. Van~Woerkom, I.~van Weperen, D.~Car, S.~R. Plissard,
	E.~P. Bakkers, L.~P. Kouwenhoven, Towards high mobility insb nanowire
	devices, Nanotechnology 26~(21) (2015) 215202.
	
	\bibitem{jalil2019new}
	A.~Jalil, S.~Agathopoulos, N.~Z. Khan, S.~A. Khan, M.~Kiani, K.~Khan, L.~Zhu,
	New physical insight in structural and electronic properties of insb
	nano-sheet being rolled up into single-wall nanotubes, Appl. Surf. Sci. 487
	(2019) 550--557.
	
	\bibitem{dong2010band}
	L.~Dong, S.~Yadav, R.~Ramprasad, S.~Alpay, Band gap tuning in gan through
	equibiaxial in-plane strains, Appl. Phys. Lett. 96~(20) (2010) 202106.
	
	\bibitem{zberecki2012emergence}
	K.~Zberecki, Emergence of magnetism in doped two-dimensional honeycomb
	structures of iii--v binary compounds, J. Supercond. Novel Magn. 25~(7)
	(2012) 2533--2537.
	
	\bibitem{xu2013stacking}
	D.~Xu, H.~He, R.~Pandey, S.~P. Karna, Stacking and electric field effects in
	atomically thin layers of gan, J. Phys.: Condens. Matter 25~(34) (2013)
	345302.
	
	\bibitem{bahuguna2016electric}
	B.~P. Bahuguna, L.~Saini, B.~Tiwari, R.~Sharma, Electric field induced
	insulator to metal transition in a buckled gaas monolayer, RCS Adv. 6~(58)
	(2016) 52920--52924.
	
	\bibitem{jalilian2016tuning}
	J.~Jalilian, M.~Naseri, S.~Safari, M.~Zarei, Tuning of the electronic and
	optical properties of single-layer indium nitride by strain and stress,
	Physica E 83 (2016) 372--377.
	
	\bibitem{zhao2016tuning}
	Q.~Zhao, Z.~Xiong, Z.~Qin, L.~Chen, N.~Wu, X.~Li, Tuning magnetism of monolayer
	gan by vacancy and nonmagnetic chemical doping, J. Phys. Chem. Solids 91
	(2016) 1--6.
	
	\bibitem{jiang2017phonon}
	Y.~Jiang, S.~Cai, Y.~Tao, Z.~Wei, K.~Bi, Y.~Chen, Phonon transport properties
	of bulk and monolayer gan from first-principles calculations, Comput. Mater.
	Sci 138 (2017) 419--425.
	
	\bibitem{zhao2015driving}
	M.~Zhao, X.~Chen, L.~Li, X.~Zhang, Driving a gaas film to a large-gap
	topological insulator by tensile strain, Sci. Rep. 5 (2015) 8441.
	
	\bibitem{raeisi2019modulated}
	M.~Raeisi, S.~Ahmadi, A.~Rajabpour, Modulated thermal conductivity of 2d
	hexagonal boron arsenide: a strain engineering study, Nanoscale 11~(45)
	(2019) 21799--21810.
	
	\bibitem{csahin2009monolayer}
	H.~{\c{S}}ahin, S.~Cahangirov, M.~Topsakal, E.~Bekaroglu, E.~Akturk, R.~T.
	Senger, S.~Ciraci, Monolayer honeycomb structures of group-iv elements and
	iii-v binary compounds: First-principles calculations, Phys. Rev. B 80~(15)
	(2009) 155453.
	
	\bibitem{zhuang2013computational}
	H.~L. Zhuang, A.~K. Singh, R.~G. Hennig, Computational discovery of
	single-layer iii-v materials, Phys. Rev. B 87~(16) (2013) 165415.
	
	\bibitem{dean2010boron}
	C.~R. Dean, A.~F. Young, I.~Meric, C.~Lee, L.~Wang, S.~Sorgenfrei, K.~Watanabe,
	T.~Taniguchi, P.~Kim, K.~L. Shepard, et~al., Boron nitride substrates for
	high-quality graphene electronics, Nat. Nanotechnol. 5~(10) (2010) 722--726.
	
	\bibitem{babaee2013substrate}
	S.~Babaee~Touski, M.~Pourfath, Substrate surface corrugation effects on the
	electronic transport in graphene nanoribbons, Appl. Phys. Lett. 103~(14)
	(2013) 143506.
	
	\bibitem{touski2020comparative}
	S.~B. Touski, M.~Hosseini, A comparative study of substrates disorder on
	mobility in the graphene nanoribbon: Charged impurity, surface optical
	phonon, surface roughness, Physica E 116 (2020) 113763.
	
	\bibitem{yao2015predicted}
	L.-Z. Yao, C.~P. Crisostomo, C.-C. Yeh, S.-M. Lai, Z.-Q. Huang, C.-H. Hsu,
	F.-C. Chuang, H.~Lin, A.~Bansil, Predicted growth of two-dimensional
	topological insulator thin films of iii-v compounds on si (111) substrate,
	Sci. Rep. 5~(1) (2015) 1--6.
	
	\bibitem{lu2018robust}
	Q.~Lu, R.~Ran, Y.~Cheng, B.~Wang, Z.-Y. Zeng, X.-R. Chen, Robust large gap
	quantum spin hall insulators in methyl and ethynyl functionalized tlsb
	buckled honeycombs, J. Appl. Phys. 124~(3) (2018) 035305.
	
	\bibitem{crisostomo2015robust}
	C.~P. Crisostomo, L.-Z. Yao, Z.-Q. Huang, C.-H. Hsu, F.-C. Chuang, H.~Lin,
	M.~A. Albao, A.~Bansil, Robust large gap two-dimensional topological
	insulators in hydrogenated iii--v buckled honeycombs, Nano Lett. 15~(10)
	(2015) 6568--6574.
	
	\bibitem{zhang2016functionalized}
	R.-w. Zhang, C.-w. Zhang, W.-x. Ji, S.-s. Li, S.-s. Yan, P.~Li, P.-j. Wang,
	Functionalized thallium antimony films as excellent candidates for large-gap
	quantum spin hall insulator, Sci. Rep. 6 (2016) 21351.
	
	\bibitem{li2016robust}
	S.-s. Li, W.-x. Ji, C.-w. Zhang, S.-j. Hu, P.~Li, P.-j. Wang, B.-m. Zhang,
	C.-l. Cao, Robust room-temperature quantum spin hall effect in
	methyl-functionalized inbi honeycomb film, Sci. Rep. 6 (2016) 23242.
	
	\bibitem{li2017gallium}
	L.~Li, O.~Leenaerts, X.~Kong, X.~Chen, M.~Zhao, F.~M. Peeters, Gallium bismuth
	halide gabi-x 2 (x= i, br, cl) monolayers with distorted hexagonal framework:
	Novel room-temperature quantum spin hall insulators, Nano Res. 10~(6) (2017)
	2168--2180.
	
	\bibitem{giannozzi2009quantum}
	P.~Giannozzi, S.~Baroni, N.~Bonini, M.~Calandra, R.~Car, C.~Cavazzoni,
	D.~Ceresoli, G.~L. Chiarotti, M.~Cococcioni, I.~Dabo, et~al., Quantum
	espresso: a modular and open-source software project for quantum simulations
	of materials, J. Phys.: Condens. Matter 21~(39) (2009) 395502.
	
	\bibitem{giannozzi2017advanced}
	P.~Giannozzi, O.~Andreussi, T.~Brumme, O.~Bunau, M.~B. Nardelli, M.~Calandra,
	R.~Car, C.~Cavazzoni, D.~Ceresoli, M.~Cococcioni, et~al., Advanced
	capabilities for materials modelling with quantum espresso, J. Phys.:
	Condens. Matter 29~(46) (2017) 465901.
	
	\bibitem{perdew1996generalized}
	J.~P. Perdew, K.~Burke, M.~Ernzerhof, Generalized gradient approximation made
	simple, Phys. Rev. Lett. 77~(18) (1996) 3865.
	
	\bibitem{goedecker1996separable}
	S.~Goedecker, M.~Teter, J.~Hutter, Separable dual-space gaussian
	pseudopotentials, Phys. Rev. B 54~(3) (1996) 1703.
	
	\bibitem{monkhorst1976special}
	H.~J. Monkhorst, J.~D. Pack, Special points for brillouin-zone integrations,
	Phys. Rev. B 13~(12) (1976) 5188.
	
	\bibitem{touski2020electrical}
	S.~B. Touski, M.~Ariapour, M.~Hosseini, Electrical and electronic properties of
	strained mono-layer inte, Physica E 118 (2020) 113875.
	
	\bibitem{bi2018thermoelectric}
	J.-Y. Bi, L.-H. Han, Q.~Wang, L.-Y. Wu, R.~Quhe, P.-F. Lu, Thermoelectric
	properties of two-dimensional hexagonal indium-va, Chin. Phys. B. 27~(2)
	(2018) 026802.
	
	\bibitem{lu2017ab}
	J.~Lu, Z.-Q. Fan, J.~Gong, X.-W. Jiang, Ab initio performance predictions of
	single-layer in--v tunnel field-effect transistors, Phys. Chem. Chem. Phys.
	19~(30) (2017) 20121--20126.
	
	\bibitem{ariapour2020strain}
	M.~Ariapour, S.~B. Touski, Strain engineering of spin and rashba splitting in
	group-iii monochalcogenide mx (m= ga, in and x= s, se, te) monolayer, J.
	Magn. Magn. Mater. (2020) 166922.
	
\end{thebibliography}

\end{document}